\documentclass[aps, pra, reprint]{revtex4-2}
\usepackage{graphicx}  
\usepackage{bm}        
\usepackage{verbatim}
\usepackage{dcolumn} 
\usepackage{amsmath} 
\usepackage{color}
\usepackage{xcolor}
\usepackage{float}
\usepackage{bbold}
\usepackage{caption}
\usepackage{subcaption}
\usepackage{booktabs}
\usepackage[justification=raggedright,singlelinecheck=false]{caption}
\newcolumntype{C}[1]{>{\centering\arraybackslash}p{#1}}

\usepackage[normalem]{ulem} 

\begin{document}
        \title{Disentanglement in dephasing channel with machine learning}
        \author{Qihang Liu}
	\email{l.qihang@wustl.edu}
	\author{Anran Qiao}
	\author{Jung-Tsung Shen}
	\affiliation{
		Department of Electrical and Systems Engineering,\\
		Washington University in St.~Louis,\\
		St.~Louis, MO 63130
	}

\begin{abstract}

Quantum state classification and entanglement quantification are of significant importance in quantum information science. Traditional methods, such as quantum state tomography, involve measurement needs that scale exponentially with the number of qubits, necessitating more efficient approaches. Recent work has shown promise in using artificial neural networks for quantum state analysis. Here we employ specialized neural networks to investigate two-qubit systems interacting with an external reservoir that induces dephasing noise, achieving high accuracy with only four features (spin correlations) and outperforming models trained on general state distributions. Our findings demonstrate the advantage of tailoring neural networks to enable efficient entanglement assessment.

\end{abstract}	

\maketitle
\pagebreak
\section{Introduction}
Quantum entanglement is essential to quantum computing, enabling efficient algorithms for certain mathematical problems that are beyond the reach of classical computing~\cite{Nielsen_Chuang_2010}. One of the most renowned examples is Shor's algorithm, which efficiently factors large integers. However, maintaining the entanglement during the quantum computing processes remains challenging~\cite{Vidal1999, Simon2002, Dur2004, Hein2005}. This challenge is primarily due to the inherent sensitivity of quantum systems to external environmental disturbances, or quantum noise, which can disrupt the delicate state of entanglement. As unwanted entanglement degradation can lead to the breakdown of quantum algorithms, accurately quantifying entanglement and categorizing quantum states become critical for quantum applications.

Traditionally, quantum state tomography is employed to measure all elements of the density operator and thoroughly characterize entanglement. In this scenario, the density operator for an $N$-qubit system has $4^N-1$ free parameters, thus the tomographic measurement demands increase exponentially with the number of qubits~\cite{Cramer2010, Chapman2016}. Nevertheless, it has also been shown that approximate reconstruction of states can be achieved with polynomial effort when a finite reconstruction error is acceptable~\cite{Lanyon2017}. Other entanglement detection methods, such as entanglement witnesses and Bell tests, require fewer measurements but can only detect a subset of entangled states~\cite{GUHNE2009}. Inspired by Bell's inequality, the neural network has been employed as a reliable quantum state classifier utilizing only a subset of tomographic features~\cite{Ma2018}. The neural network can be regarded as a combination of various Bell-like inequalities, constructing a nonlinear function to represent the boundary between entangled and separable states. Compared with the Bell test, the neural network approach provides a sufficient and necessary method. Recently, machine-learning algorithms have been applied to quantum systems~\cite{Lloyd2014, Biamonte2017, Maria2015, Torlai2018}. In particular, artificial neural networks (ANNs) have demonstrated efficient and reliable state classification~\cite{Ma2018} and entanglement quantification~\cite{Dominik2023}. 

Previous studies have employed neural networks to analyze the effects of depolarizing noise. Nonetheless, the most commonly encountered type of noise in quantum computing is dephasing noise~\cite{albash2015}. In this paper, we apply machine-learning methods to investigate the effect of dephasing noise on the disentanglement process. We introduce ANN algorithms specifically trained to classify quantum states and quantify entanglement affected by dephasing noise. Unlike quantum state tomography, which requires measuring all 15 spin correlations, our machine-learning-assisted approach achieves an accuracy of 86.9\% using only four spin correlations. For multi-qubit systems, machine-learning algorithms are expected to offer even greater advantages over quantum state tomography.

We trained a specialized neural network on datasets comprising states affected by the dephasing noise. Our findings reveal that specialized ANNs significantly outperform models trained on other data distributions. In the task of classifying states under dephasing noise, these specialized models achieve notably higher accuracy.

This research demonstrates the potential of machine learning in the two-qubit disentanglement process and paves the way for more efficient approaches to quantum state classification and entanglement assessment in multi-qubit systems.

\section{Disentanglement process in dephasing channel}
In preparation for the machine learning models, this section discusses the Kraus representation of the dephasing channel and derives the evolution of a two-qubit density operator during the dephasing process. We employ the concept of concurrence as a measure of the degree of entanglement and discuss the concurrence properties for a dephasing two-qubit state.




\subsection{Kraus representation of dephasing channel}
The dephasing channel, also known as the phase-damping channel, is a ubiquitous noise process where partial quantum information is lost without any loss of energy. Physically, the interaction between qubits and their environment leads to entanglement among all components, resulting in information loss when the environmental states are traced out. For instance, the atom as a qubit can experience the dephasing process by interacting with a stochastic far-off-resonant electromagnetic wave. To model the dephasing process, the Kraus operators are generally used~\cite{Caruso2014}. 


To begin with, we discuss the Kraus operators for the simplest case of single qubits. The dephasing process during a short time interval can be expressed by the following,
\begin{align}
    \hat{O}_D(p)|0\rangle_A|0\rangle_E &= \sqrt{1-p}|0\rangle_A|0\rangle_E+\sqrt{p}|0\rangle_A|1\rangle_E\notag\\
    \hat{O}_D(p)|1\rangle_A|0\rangle_E &= \sqrt{1-p}|1\rangle_A|0\rangle_E+\sqrt{p}|1\rangle_A|2\rangle_E,
\end{align}
where $\hat{O}_D(p)$ is a unitary time evolution operator on the system containing the qubit and its surrounding environment. Here, $p$ is the noise strength, denoting the probability of a scattering event between the qubit and the environment state. 

By tracing out the environment states, the qubit-state evolution in the dephasing process can be expressed as $\hat{\rho}_A'= \sum_{a=1,2,3}\hat{M}_a\hat{\rho}_A\hat{M}_a$, where $\hat{\rho}_A$ and $\hat{\rho}_A'$ are the density matrix of the qubit before and after the dephasing noise. $\hat{M}_a$ ($a=1,2,3$) represents the Kraus operators,
\begin{align}
    &\hat{M}_1 = \left(\begin{array}{cc}
        \sqrt{1-p} & 0 \\
        0 & \sqrt{1-p}
    \end{array}\right), \notag\\
    &\hat{M}_2 = \left(\begin{array}{cc}
        \sqrt{p} & 0 \\
        0 & 0
    \end{array}\right), \quad\hat{M}_3 = \left(\begin{array}{cc}
        0 & 0 \\
        0 & \sqrt{p}
    \end{array}\right).
\end{align}

The effect of dephasing channel on a single qubit is the loss of relative phases between $|0\rangle$ and $|1\rangle$, which can be expressed as,
\begin{align}
    \hat{\rho}_A' = \left(\begin{array}{cc}
        \rho_{00} & \rho_{01}(1-p) \\
        \rho_{10}(1-p) & \rho_{11}
    \end{array}\right),
\end{align}
where the density operator for the two-qubit state before dephasing is $\hat{\rho}_A = \{\rho_{ij}\}_{0\leq i\leq 1, 0\leq j\leq1}$.

For the two-qubit states, the dephasing process can be express as the following,
\begin{widetext}
\begin{align}       
    \hat{O}_D(p_1)\otimes\hat{O}_D(p_2)|i\rangle_{A_1}|j\rangle_{A_2}|0\rangle_{E_1}|0\rangle_{E_2} &= \left(\sqrt{1-p_1}|i\rangle_{A_1}|0\rangle_{E_1}+\sqrt{p_1}|i\rangle_{A_1}|i+1\rangle_{E_1}\right)\left(\sqrt{1-p_2}|j\rangle_{A_2}|0\rangle_{E_2}+\sqrt{p_2}|j\rangle_{A_2}|j+1\rangle_{E_2}\right),
\end{align}
where $p_1$ and $p_2$ are the noise strengths for each qubits separately. Hereafter, we assume the global dephasing channel scenario, where $p_1=p_2=p$~\cite{Yu2003, Braun2002, Yu2002}. After dephasing noise, the two-qubit density matrix $\hat{\rho}_A=\{\rho_{ij}\}_{0\leq i\leq 3, 0\leq j\leq3}$ evolves into $\hat{\rho}_A'=\sum_{i,j=1,2,3}\hat{\mathcal{E}}_{ij}\hat{\rho}_A\hat{\mathcal{E}}_{ij}$, where the Kraus operators $\mathcal{E}_{ij}$ have the following form, 
\begin{align}
    &\hat{\mathcal{E}}_{11} =  \begin{pmatrix}
    1-p & 0 & 0 & 0 \\
    0 & 1-p & 0 & 0 \\
    0 & 0 & 1-p & 0 \\
    0 & 0 & 0 & 1-p
    \end{pmatrix}, \quad
    \hat{\mathcal{E}}_{12} =  \begin{pmatrix}
    \sqrt{p(1-p)} & 0 & 0 & 0 \\
    0 & 0 & 0 & 0 \\
    0 & 0 & \sqrt{p(1-p)} & 0 \\
    0 & 0 & 0 & 0
    \end{pmatrix}, \quad
    \hat{\mathcal{E}}_{13} = \begin{pmatrix}
    0 & 0 & 0 & 0 \\
    0 & \sqrt{p(1-p)} & 0 & 0 \\
    0 & 0 & 0 & 0 \\
    0 & 0 & 0 & \sqrt{p(1-p)}
    \end{pmatrix}, \notag \\
    &\hat{\mathcal{E}}_{21} =  \begin{pmatrix}
    \sqrt{p(1-p)} & 0 & 0 & 0 \\
    0 & \sqrt{p(1-p)} & 0 & 0 \\
    0 & 0 & 0 & 0 \\
    0 & 0 & 0 & 0
    \end{pmatrix}, \quad
    \hat{\mathcal{E}}_{22} =  \begin{pmatrix}
    p & 0 & 0 & 0 \\
    0 & 0 & 0 & 0 \\
    0 & 0 & 0 & 0 \\
    0 & 0 & 0 & 0
    \end{pmatrix}, \quad
    \hat{\mathcal{E}}_{23} =  \begin{pmatrix}
    0 & 0 & 0 & 0 \\
    0 & p & 0 & 0 \\
    0 & 0 & 0 & 0 \\
    0 & 0 & 0 & 0
    \end{pmatrix}, \notag \\
    &\hat{\mathcal{E}}_{31} =  \begin{pmatrix}
    0 & 0 & 0 & 0 \\
    0 & 0 & 0 & 0 \\
    0 & 0 & \sqrt{p(1-p)} & 0 \\
    0 & 0 & 0 & \sqrt{p(1-p)}
    \end{pmatrix}, \quad
    \hat{\mathcal{E}}_{32} =  \begin{pmatrix}
    0 & 0 & 0 & 0 \\
    0 & 0 & 0 & 0 \\
    0 & 0 & p & 0 \\
    0 & 0 & 0 & 0
    \end{pmatrix}, \quad
    \hat{\mathcal{E}}_{33} =  \begin{pmatrix}
    0 & 0 & 0 & 0 \\
    0 & 0 & 0 & 0 \\
    0 & 0 & 0 & 0 \\
    0 & 0 & 0 & p
    \end{pmatrix}.\label{Eq:Kraus_operators}
\end{align}
$\hat{\mathcal{E}}_{ij}$ is the tensor products of single-qubit Kraus operator $\hat{M}_a$. Therefore,
\begin{align}
    \hat{\rho}_A' = \left(\begin{array}{cccc}
        \rho_{00} & (1-p)\rho_{01} & (1-p)\rho_{02} & (1-p)^2\rho_{03} \\
        (1-p)\rho_{10} & \rho_{11} & (1-p)^2\rho_{12} & (1-p)\rho_{13} \\
        (1-p)\rho_{20} & (1-p)^2\rho_{21} & \rho_{22} & (1-p)\rho_{23} \\
        (1-p)^2\rho_{30} & (1-p)\rho_{31} & (1-p)\rho_{32} & \rho_{33} 
    \end{array}\right). \label{global_density_operator}
\end{align}
\end{widetext}
In continuous dephasing processes, the influence of dephasing can be expressed by replacing $1-p$ with $e^{-\Gamma t}$, where $\Gamma$ is the scattering probability per unit time. In both one- and two-qubit dephasing processes, the off-diagonal elements in the density operators decay exponentially, representing the loss of relative phases.


\subsection{Entanglement properties of two-qubit states via dephasing channel}
\label{subsec:analytical_concurrence_dephasing}

The degree of entanglement is quantified by concurrence~\cite{Hill1997, Wootters1998}. The concurrence emerges in the derivation of entanglement of formation when it's expressed as an explicit function of the density operator. For a general two-qubit state with density operator $\hat{\rho}$, the concurrence is defined as
\begin{align}
    C(\hat{\rho}) = \max\{0,\sqrt{\lambda_1}-\sqrt{\lambda_2}-\sqrt{\lambda_3}-\sqrt{\lambda_4}\}, \label{concurrence}
\end{align}
where $\lambda_i$ ($i=1,2,3,4$) are eigenvalues of the operator $\hat{\rho}(\hat\sigma_y\otimes\hat\sigma_y)\hat{\rho}^*(\hat\sigma_y\otimes\hat\sigma_y)$ in the decreasing order. Here $\hat\sigma_y$ is the Pauli matrix, and the complex conjugate $\hat\rho^*$ is taken in the standard basis.  The concurrence ranges from 0 to 1, where 0 represents separable states and 1 represents maximally entangled states, such as Bell states.

A general two-qubit pure state can be expanded with four computational basis, and can be expressed as follows:
\begin{align}
    |\psi\rangle = c_1|00\rangle + c_2|01\rangle + c_3|10\rangle + c_4|11\rangle, \label{general_purestate}
\end{align}
for which, the concurrence can be shown to be
\begin{align}
    C(\psi) = 2\left|c_1c_4-c_2c_3\right|.
\end{align}
It's worth noting that for any two-qubit pure states expanded with three or fewer computational basis, the classification of entanglement remains unchanged in the global dephasing channel~\cite{Yu2003}. For instance, for the following state,
\begin{align}
    |\psi\rangle = c_1|00\rangle + c_2|01\rangle + c_3|10\rangle.
\end{align}
Before dephasing noise is applied, the concurrence is $2|c_2c_3|$. After experiencing the global dephasing noise, the concurrence becomes $2(1-p)^2|c_2c_3|$. For an initially separable pure state, the concurrence remains zero, and the state stays separable throughout the dephasing process. Conversely, for an initially entangled state, the state remains entangled, but its degree of entanglement decreases exponentially over time, as the dynamics of dephasing can be expressed by replacing $1-p$ to $e^{-\Gamma t}$. Therefore, only pure states in Eq.~\ref{general_purestate} with four non-zero amplitudes can experience transition from entangled to separable, vice versa. 


In the following, we introduce the state classification and entanglement quantification models to predict the entanglement properties of the two-qubit system subjected to dephasing noise.  

\begin{figure}
    \centering
    \includegraphics[width=.5\textwidth]{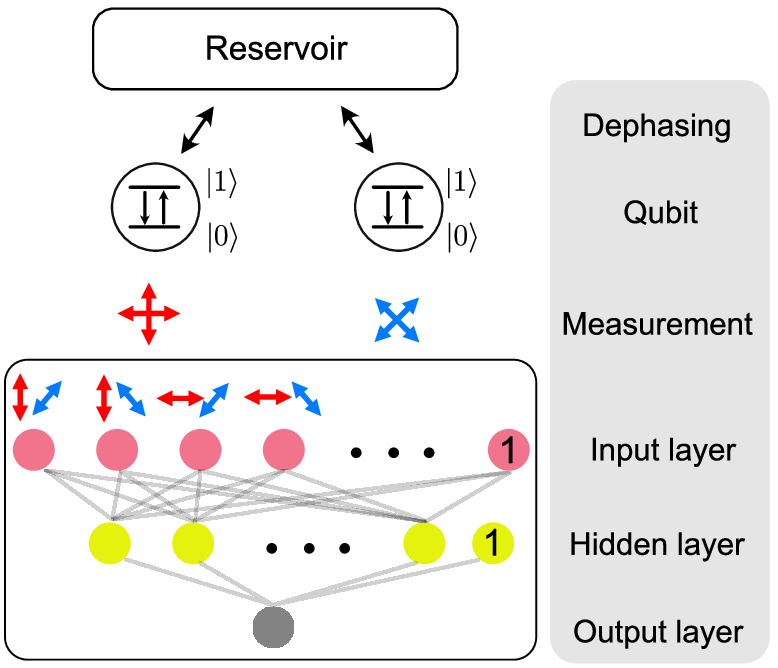}
    \caption{Schematic representation of a two‐qubit system interacting with an external reservoir that induces dephasing. The measured spin correlations (red and blue arrows) serve as inputs to the ANN, which consists of an input layer (pink nodes), a single hidden layer (yellow nodes), and an output layer (gray node). The nodes labeled “1” represent the bias terms.}
    \label{fig:schematics}
\end{figure}

\section{Entanglement Classification Model}
\label{sec:Classification}
To begin with, we construct the datasets as follows: we form an ensemble of two-qubit states by randomly distributed parameters characterizing the state, followed by the formation of the density operators. Each entry of the dataset comprises fifteen tomographic features; the first nine features encapsulate all spin correlations $\hat{\sigma}_i\otimes\hat{\sigma}_j$ between the qubits, while the remaining six features detail the spin values of individual qubits.

The features are directly computed from the density operator of the state. Similar to the setup in the typical CHSH experiment, the spins are measured in different directions~\cite{CHSH1969}. For the first qubit, the spin is measured in the \{$x, y, z$\} direction. The measurement direction for the second qubit is defined by the set \{$a, b, z$\}, which represents a coordinate system obtained by rotating the \{$x, y, z$\} coordinate system 45 degrees anticlockwise around the z-axis, as shown in FIG.~\ref{fig:schematics}. The transformation of the Pauli spin matrices for the second qubit under this rotation is given by:
\begin{align}
    \hat\sigma_a &= \hat\sigma_x\cos\frac{\pi}{4}+\hat\sigma_y\sin\frac{\pi}{4}\notag\\
    \hat\sigma_b &= -\hat\sigma_x\sin\frac{\pi}{4}+\hat\sigma_y\cos\frac{\pi}{4}\notag\\
    \hat\sigma_z &= \hat\sigma_z.
\end{align}

\begin{table*}
\caption{\label{tab:order}Order of features in each entry}
\begin{ruledtabular}
\begin{tabular}{cccccccccc}
 \textbf{Index}& 1 &2&3&4&5&6&7&8&9\\
 \textbf{Feature} &$\langle\hat\sigma_x\otimes\hat\sigma_a\rangle$ &$\langle\hat\sigma_x\otimes\hat\sigma_b\rangle$&   $\langle\hat\sigma_y\otimes\hat\sigma_a\rangle$&$\langle\hat\sigma_y\otimes\hat\sigma_b\rangle$&$\langle\hat\sigma_x\otimes\hat\sigma_z\rangle$&$\langle\hat\sigma_y\otimes\hat\sigma_z\rangle$&$\langle\hat\sigma_z\otimes\hat\sigma_a\rangle$&$\langle\hat\sigma_z\otimes\hat\sigma_b\rangle$&$\langle\hat\sigma_z\otimes\hat\sigma_z\rangle$\\
 \hline
  \textbf{Index}&10&11&12&13&14&15\\
 \textbf{Feature} &$\langle\hat\sigma_x\otimes \hat{I}\rangle$ &$\langle\hat\sigma_y\otimes \hat I\rangle$&$\langle\hat\sigma_z\otimes\hat I\rangle$&$\langle\hat I\otimes\hat\sigma_a\rangle$&$\langle\hat I\otimes\hat\sigma_b\rangle$&$\langle\hat I\otimes\hat\sigma_z\rangle$\\
\end{tabular}
\end{ruledtabular}
\end{table*}
The features are arranged in the order as shown in Table~\ref{tab:order}. The first four features represent the spin correlations in CHSH inequality (FIG.~\ref{fig:schematics}). Each data is labeled by applying the Peres–Horodecki positive partial transpose (PPT) separability criterion~\cite{Horodecki1996}, where the entangled state is labeled ``1" and the separable state is labeled ``0".

The neural network is trained using different subsets of features. Specifically, we discuss the models trained using only the first four features (referred to as the Bell-like classifier) and the complete set of fifteen features (termed the tomographic classifier). Upon completion of the training, the neural network is used to predict the entanglement classification of an unknown quantum state, utilizing the measured spin correlations and the values of individual spins.

\begin{figure*}[htbp]
    \centering
    \includegraphics[width=.9\textwidth]{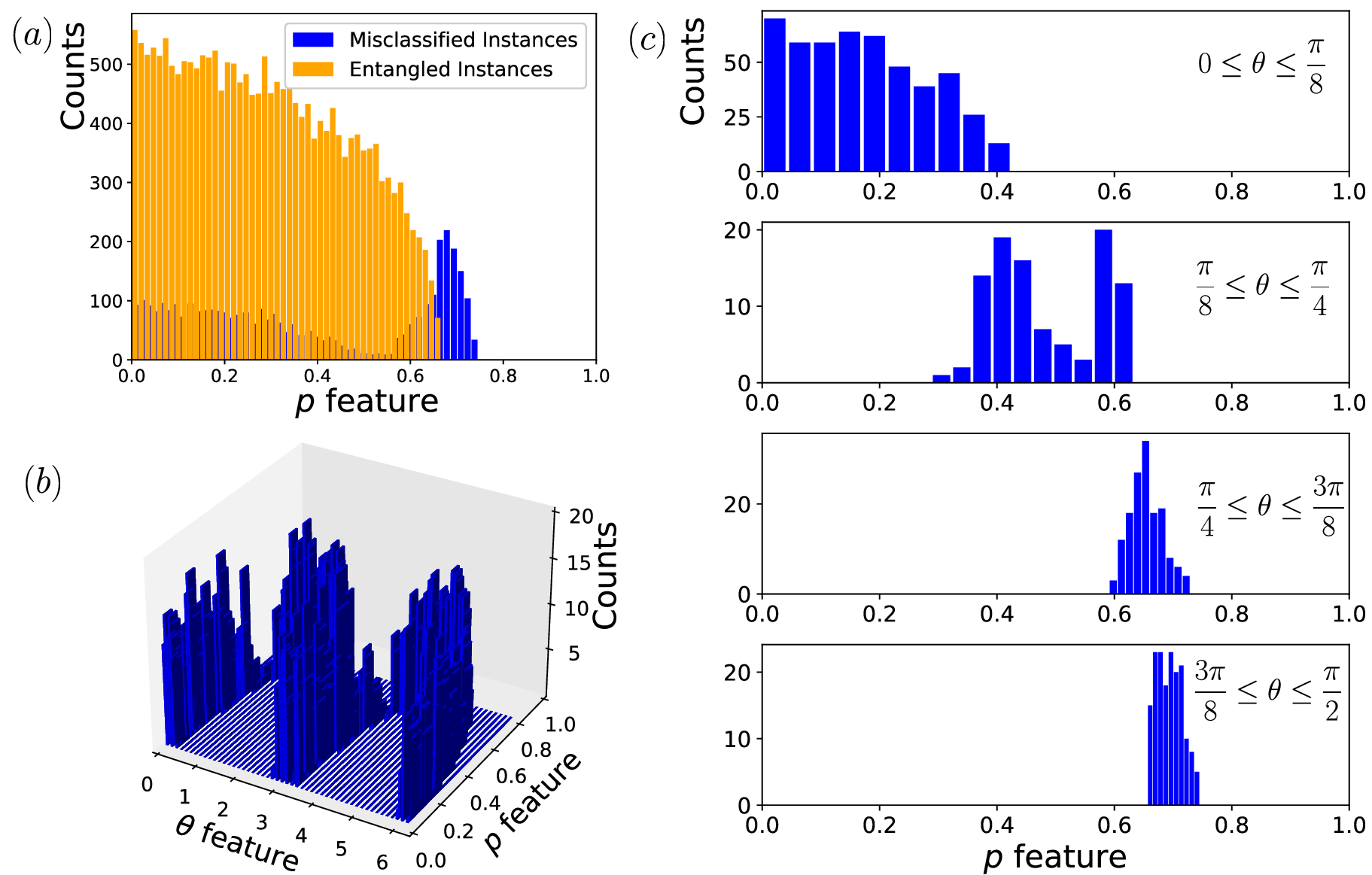} 
    \caption{Distribution of instances for the Bell-like classifier in the depolarizing model. (a) Distribution of misclassified (blue) and entangled (orange) instances over $p$ parameter. (b) Distribution of misclassified instances over $p$ and $\theta$ parameters. (c) Distribution of misclassified instances over $p$ parameter in different $\theta$ intervals ($\theta\in$$[0,\pi/8]$, $[\pi/8,\pi/4]$, $[\pi/4,3\pi/8]$, and $[3\pi/8,\pi/2]$).}
    \label{fig_depolarizing}
\end{figure*}
\subsection{Depolarizing channel}
\label{subsec:DepolarizingChannel}
As a test of our neural network model, we first look into a special family of two-qubit entangled pure states experiencing depolarizing noise, which has been studied in Ref.~\cite{Ma2018}. The special family of states can be expressed as,
\begin{align}
    |\psi_1\rangle = \cos\frac{\theta}{2}|00\rangle+\sin\frac{\theta}{2}e^{i\varphi}|11\rangle,
\end{align}
where both $\theta$, $\varphi$ parameters are randomly distributed from 0 to $2\pi$. After the depolarizing noise, the density operator of the states become 
\begin{align}
    \hat{\rho} = (1-p)|\psi_1\rangle \langle\psi_1| + \frac{p}{4}\hat{I},
\end{align}
where $p$ is the noise strength in the depolarizing channel. In the dataset, the noise strength $p$ is randomly distributed from 0 to 1. Applying the PPT criterion, the states remain entangled only when $\frac{p}{4}-(1-p)\cos\frac{\theta}{2}\sin\frac{\theta}{2}<0$. 


\textbf{Model}: The model is developed using the Keras framework. It consists of a single hidden layer with 15 nodes and employs the ReLU activation function. The network is compiled using the binary cross entropy loss function and the stochastic gradient descent (SGD) optimization method. We set the batch size at $n_{\text{batch}}=32$ and train over 200 epochs. The learning rate in the SGD optimizer is set as 0.01. 

A dataset comprising 200,000 entries (80\% designated for training and 20\% for testing) is utilized to train the neural network. The dataset size has been shown sufficient for the classification of states experiencing the depolarizing channel~\cite{Ma2018}. For the present dataset size, We found that more nodes in the hidden layer will not significantly improve the model performance. For the Bell-like classifier, the distribution of misclassified and entangled instances across varying strengths of depolarizing noise ($p$) is depicted in FIG.~\ref{fig_depolarizing}(a). It is observed that the number of entangled states diminishes with increasing noise strength. Analytical results indicate that all states become separable when the noise strength exceeds $\frac{2}{3}$. Notably, a peak in the number of misclassified instances occurs at this boundary ($p = \frac{2}{3}$) between entangled and separable states.



The distribution of misclassified instances over varying noise strengths and the $\theta$ parameter is illustrated in FIG~\ref{fig_depolarizing}(b,c). This distribution is further detailed by the histograms, which are segmented into intervals of the $\theta$ parameter: $[0,\pi/8]$, $[\pi/8,\pi/4]$, $[\pi/4,3\pi/8]$, and $[3\pi/8,\pi/2]$. It is observed that misclassified instances predominantly occur at the boundary between entangled and separable states, defined by the equation $\frac{p}{4}-(1-p)\cos\frac{\theta}{2}\sin\frac{\theta}{2}=0$. The observed distribution of misclassified instances aligns closely with the findings reported in Ref.~\cite{Ma2018}, consolidating the reliability of the machine learning method for entanglement classification.

\label{subsec:Dephasing}
\begin{figure*}[htbp]
    \centering
    \includegraphics[width=.9\textwidth]{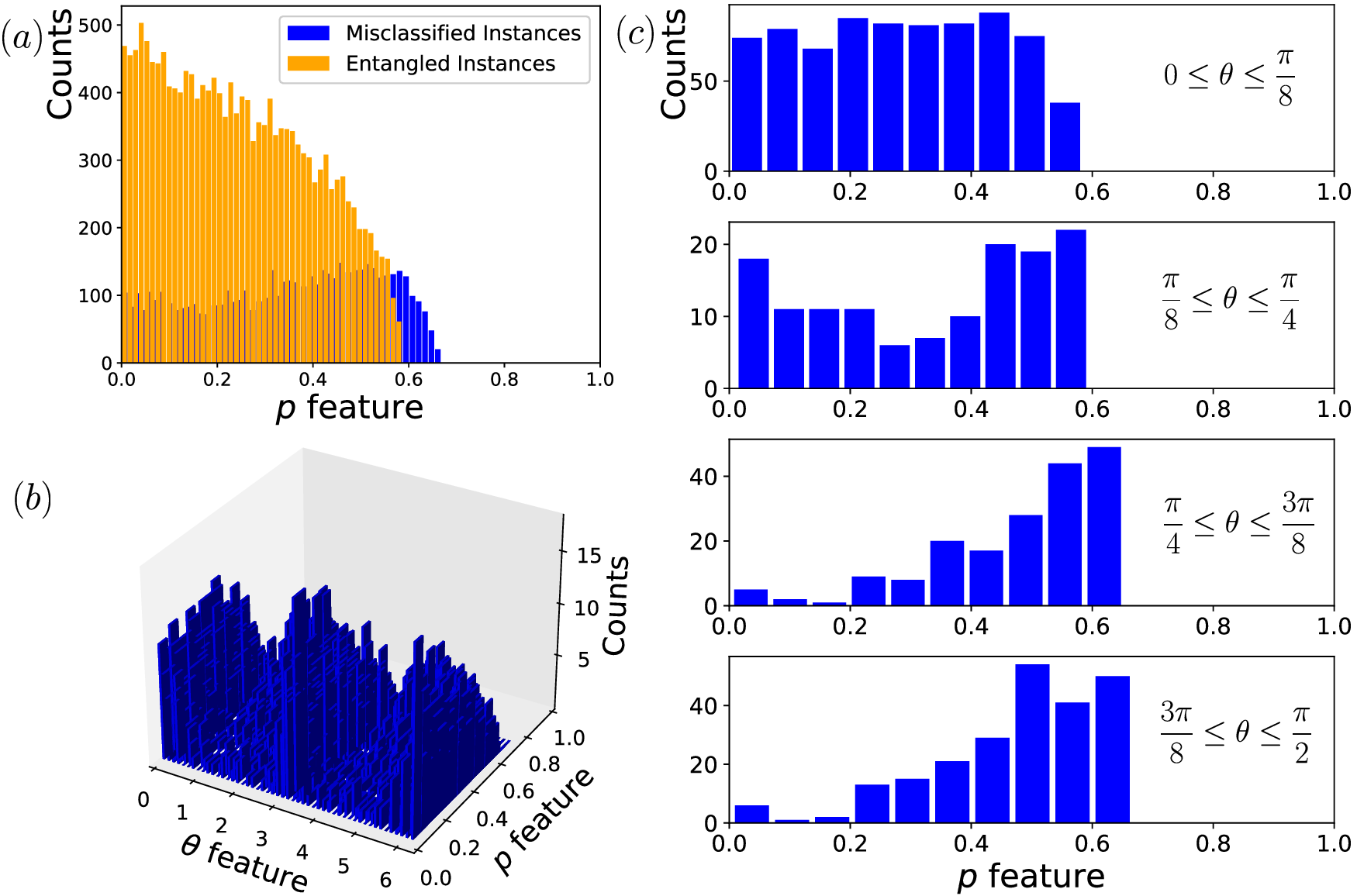}
    \caption{Distribution of instances for the Bell-like classifier in the dephasing model. (a) Distribution of misclassified (blue) and entangled (orange) instances over $p$ parameter. (b) Distribution of misclassified instances over $p$ and $\theta$ parameters. (c) Distribution of misclassified instances over $p$ parameter in different $\theta$ intervals ($\theta\in$$[0,\pi/8]$, $[\pi/8,\pi/4]$, $[\pi/4,3\pi/8]$, and $[3\pi/8,\pi/2]$).}
    \label{fig_dephasing}
\end{figure*}
\subsection{Dephasing channel}
Building upon the positive outcomes observed in the depolarizing channel, we investigate the dephasing channel. This section focuses on the disentanglement process for a specific family of pure states characterized by four non-zero components, which can transition from entangled to separable states. These states are represented as:
\begin{align}
    |\psi_2\rangle = \cos\frac{\theta}{2}|0\rangle\left(\frac{e^{-i\varphi}|0\rangle+|1\rangle}{\sqrt{2}}\right)+\sin\frac{\theta}{2}|1\rangle\left(\frac{e^{i\varphi}|0\rangle+|1\rangle}{\sqrt{2}}\right),
\end{align}
where both $\theta$, $\varphi$ parameters range from 0 to $2\pi$. This special family of states is generated by applying controlled Hadamard operations on the $|\psi_1\rangle$ states. These states are pertinent in quantum teleportation and quantum information processing~\cite{Nielsen_Chuang_2010}. After the dephasing noise, the density operator for the states can be expressed as $\hat\rho_2=\sum_{i,j=1,2,3}\hat{\mathcal{E}}_{ij}|\psi_2\rangle\langle\psi_2|\hat{\mathcal{E}}_{ij}$, where $\hat{\mathcal{E}}_{ij}$ are the Kraus operators for the dephasing channel in Eq.~\ref{Eq:Kraus_operators}. The dephasing noise strength $p$ in the Kraus operators is randomly distributed from 0 to 1. The entanglement ratio in the generated dataset is approximately $42\%$.

The neural network model designated for the dephasing channel is trained utilizing the identical architecture and parameters described in Section~\ref{subsec:DepolarizingChannel}. 

The distribution of misclassified and entangled instances under dephasing noise is depicted in FIG.~\ref{fig_dephasing}(a).  It is noted that the quantity of entangled instances declines as the dephasing noise strength increases, with all states becoming separable when the noise exceeds a strength of $p > 0.6$. Notably, a minor peak in misclassified instances occurs at the entanglement boundary. In the physical processes, such as superconducting qubits affected by the fluctuations of external electromagnetic fields, the noise strength $p$ can be replaced by $(1-e^{-\Gamma t})$. During time evolution, the influence of dephasing noise on the qubits manifests as an increasing $p$ parameter in the density operators, starting from $p=0$ for initial states and approaching $p=1$ after sufficiently long times. The accuracy of the machine learning model is relatively lower within a time range where the states lie near the boundary between entangled and separable states. Additionally, we have illustrated the distribution of misclassified instances across variations in noise strength and the $\theta$ parameter in FIG.~\ref{fig_dephasing}(b), as well as across different $\theta$ intervals in FIG.~\ref{fig_dephasing}(c). An increased occurrence of misclassification is observed when $\theta$ approaches $0$, $\pi$, and $2\pi$. These distributions elucidate the boundaries between entangled and separable states. 


\subsection{General two-qubit mixed states}
\label{subsec:GT_data_generation}
\begin{figure}[htbp]
    \centering
    \includegraphics[width=.5\textwidth]{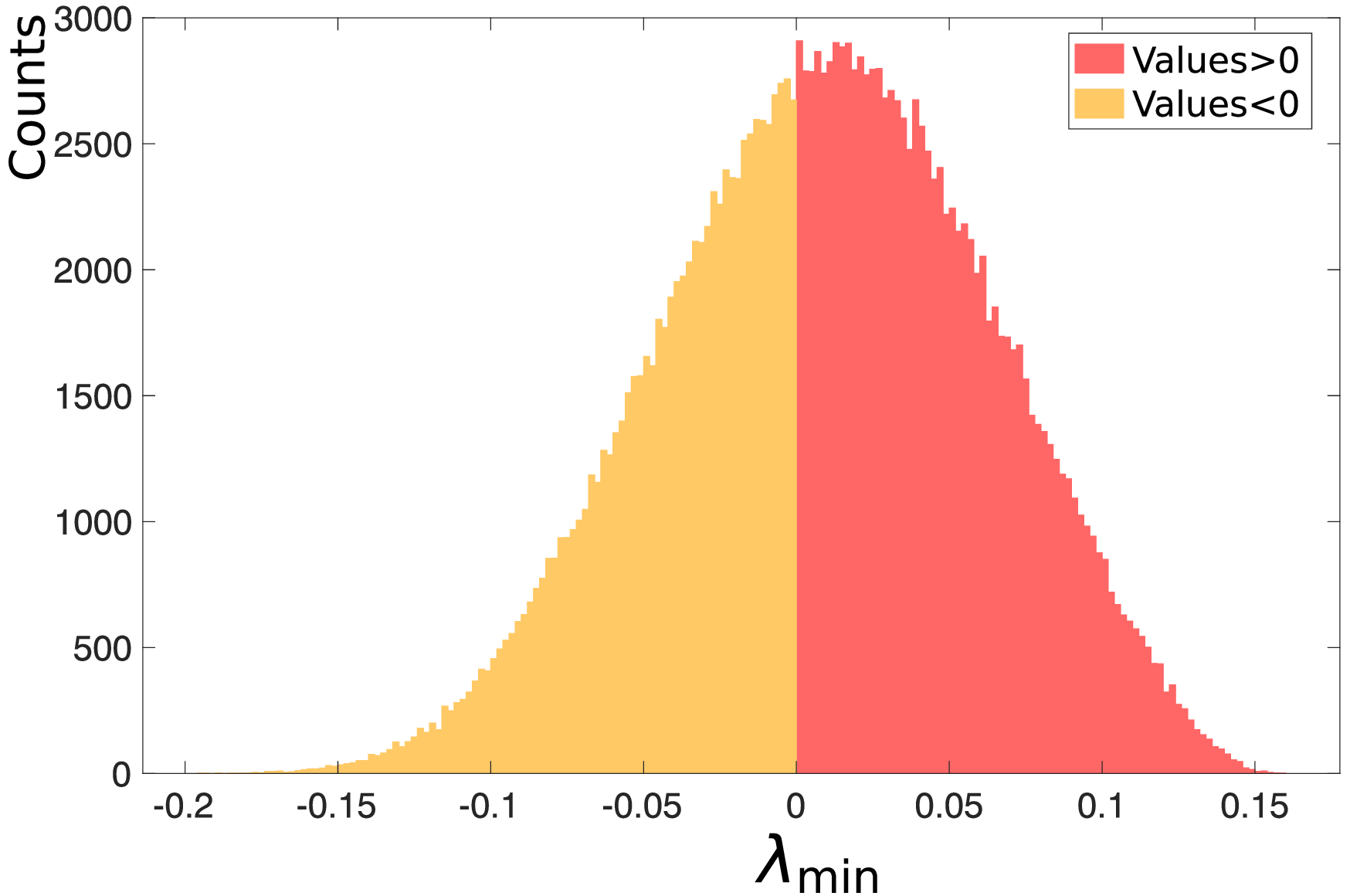}
    \caption{Distribution of the minimum eigenvalues of the partially transposed density matrices for the general two-qubit mixed states dataset. Entangled states are represented in yellow, while separable states are represented in red.}
    \label{fig:min-eigen value}
\end{figure}
For the general two-qubit mixed state, the dataset is constructed with the aid of QR decomposition to generate random $4\times4$ matrices serving as density operators $\hat{\rho}=\hat{U}\hat{\Lambda}\hat{U}^\dagger$~\cite{trefethen2022}. The randomness is introduced by varying the eigenvalues of the diagonal matrix $\hat{\Lambda}$, and employing random unitary transformation matrices $\hat{U}$.



The distribution of entangled states within this dataset is depicted through the minimum eigenvalues $\lambda_{min}$ of the partially transposed density matrices, as illustrated in FIG.~\ref{fig:min-eigen value}. According to the PPT criterion, states with negative $\lambda_{min}$ are entangled, while those with positive $\lambda_{min}$ values are separable. To ensure data consistency across different datasets, we adjust the distribution of eigenvalues in the $\Lambda$ matrices, setting the entanglement ratio at approximately 42\%, as in Section~\ref{sec:Classification}B.
\begin{figure*}[htbp]
    \centering
    \includegraphics[width=.8\textwidth]{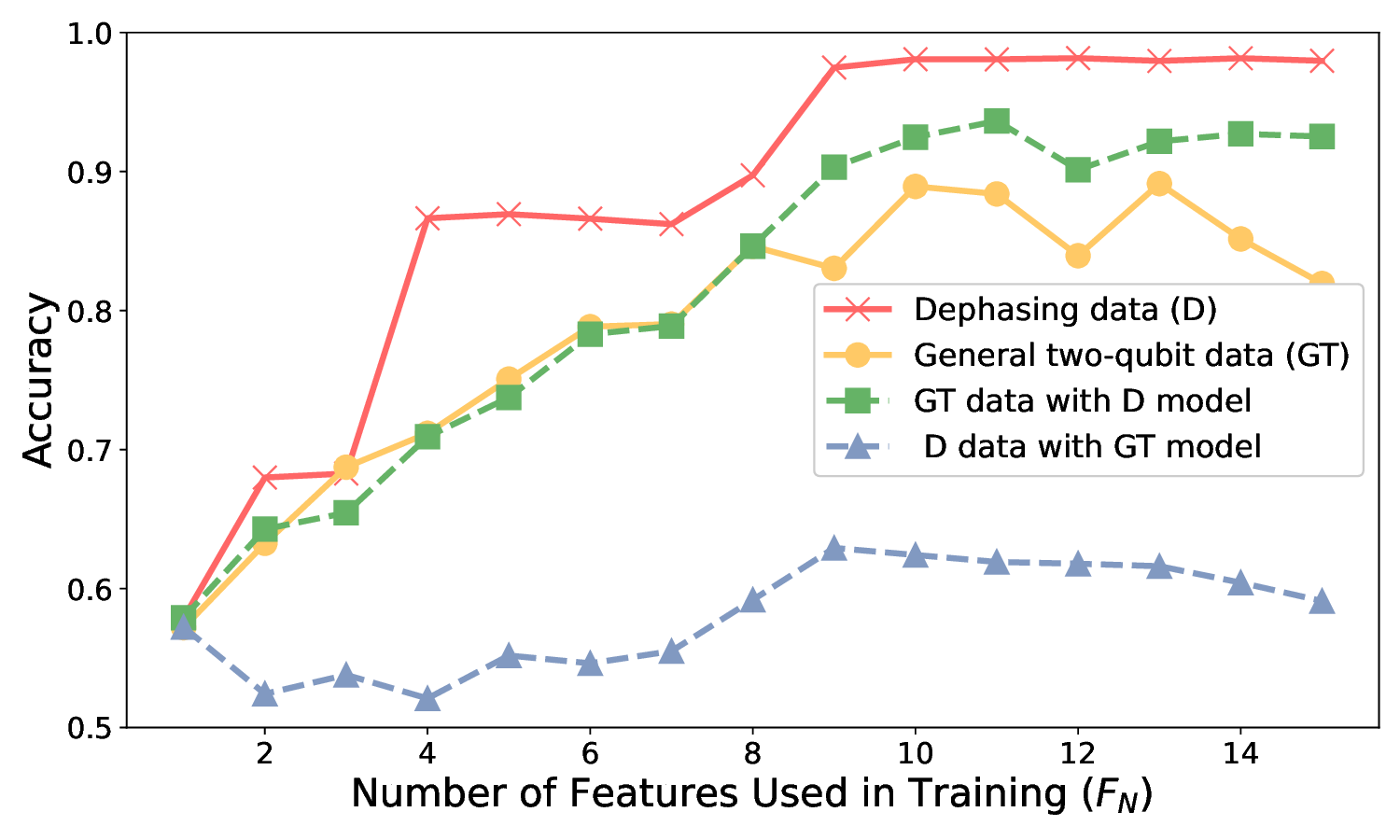}
    \caption{Model performance (accuracy) as a function of feature number for different datasets. The red line represents models trained with special 4-basis pure states subjected to dephasing noise. The yellow line indicates models trained on general two-qubit data. The green line depicts the performance of models initially trained on dephasing noise data and subsequently applied to general two-qubit data. Conversely, the blue line shows the performance for models trained on general two-qubit data tested against dephasing noise datasets.}
    \label{fig:4_way_comparasion}
\end{figure*}
\subsection{Classification model performance comparison}

\begin{figure*}[htbp]
    \centering
    \includegraphics[width=0.8\textwidth]{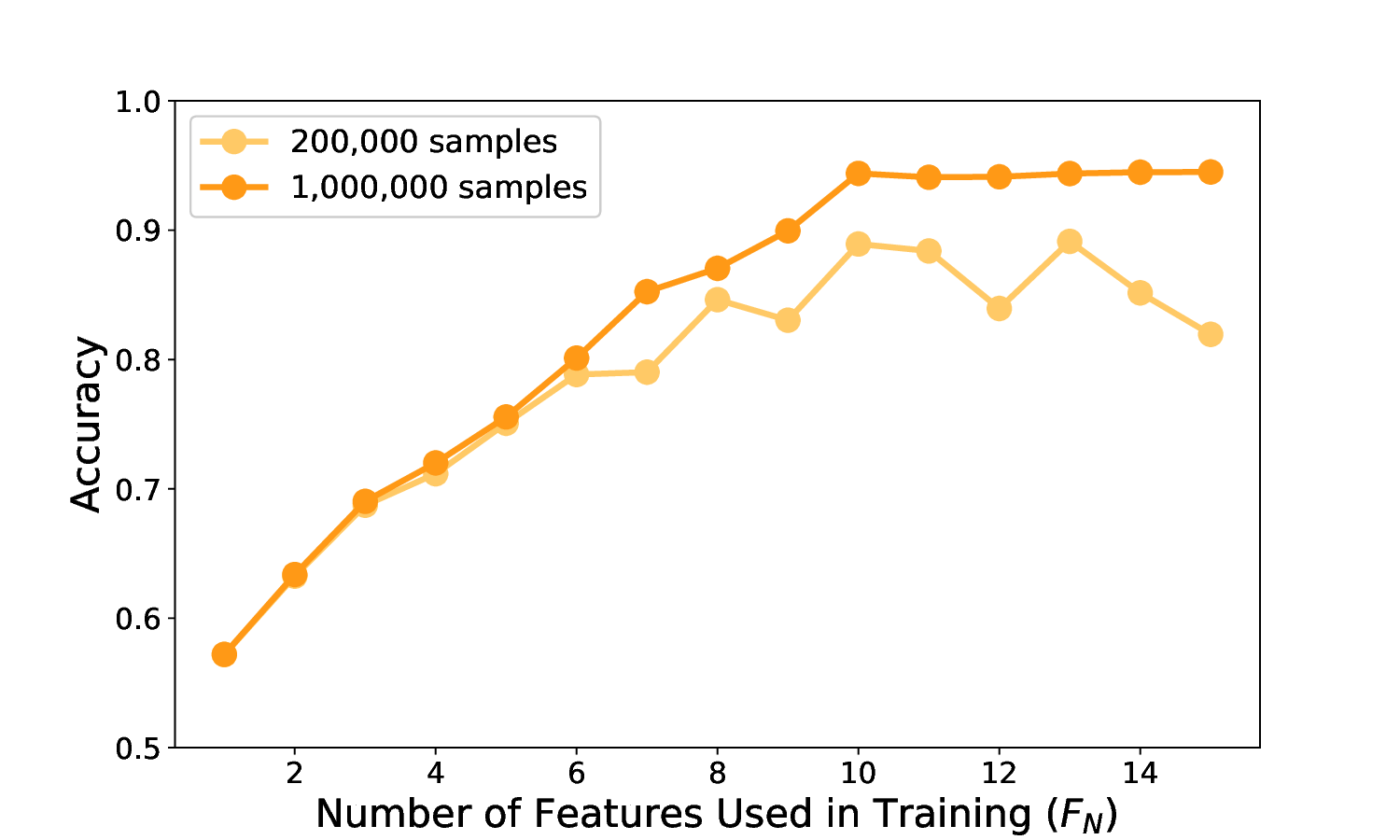}
    \caption{Model performance (accuracy) as a function of feature number for general two-qubit mixed state datasets across different model complexities and dataset sizes. The yellow line represents the model trained with 15 nodes in one hidden layer and 200,000 samples. The orange line represents the model trained with 100 nodes in one hidden layer and 1,000,000 samples.}
    \label{fig:GT_comparasion}
\end{figure*}

Here we compare the performance of the dephasing model (D) and general two-qubit model (GT). FIG.~\ref{fig:4_way_comparasion} plots the accuracy of the two models, trained on different dataset, as a function of the feature numbers. The features are organized according to the order presented in Table~\ref{tab:order}. Our result shows that the dephasing model (red line) with all tomographic features can achieve a remarkable accuracy of 98.0\%. In contrast, the general two-qubit model (yellow line) using the same set of features can only achieve a moderate accuracy of 81.9\%.

In the dephasing model, two plateaus are observed when number of features used in training are four and nine, respectively. The plateaus underscores the importance of the first four features (Bell-like classifier) and all nine spin correlations in the entanglement classification. For the general two-qubit model, performance progressively increases with the addition of more features but exhibits fluctuations when more than nine features are used. These fluctuations are attributed to underfitting, a consequence of the dataset's complexity and the simplicity of the neural network structure. Despite identical dataset sizes and model settings, the dephasing noise model significantly outperforms the general two-qubit mixed states model. Specifically, the Bell-like classifiers for dephasing dataset and general two-qubit dataset achieve accuracies of 86.6\%, and 71.2\%, respectively. 

To assess the adaptability of the models, we also conduct the cross-comparisons between the dephasing and general two-qubit datasets. As shown in FIG.~\ref{fig:4_way_comparasion}, general two-qubit data with dephasing model (green line) with tomographic features can achieve a high accuracy of 92.5\%. In contrast, dephasing data with general two-qubit model (blue) using the same set of features achieves a low accuracy of 59.1\%, which is only marginally better than random guessing. The low accuracy again arises from the underfitting of the general two-qubit model.

The density matrix in the dephasing dataset contains three parameters, and is the subset of those in the general two-qubit dataset, which has 15 parameters. With equivalent model complexities, the dephasing model can capture the entanglement properties more effectively due to the simplicity of the dataset. 

The performance of the general two-qubit model can be significantly improved by increasing both the dataset size and the model complexity, as shown in FIG.~\ref{fig:GT_comparasion}. A comparison between the model trained with 15 nodes in a single hidden layer using 200,000 samples and the model trained with 100 nodes in a single hidden layer using 1,000,000 samples demonstrates a significant improvement across various feature subsets. In both cases, 80\% of samples are used for training. As expected, the fluctuations observed in FIG.~\ref{fig:4_way_comparasion} disappeared with increased model complexity and larger datasets. For the later case, the accuracy increases monotonically up to $F_n = 9$, stabilizing into a performance plateau with more than nine features. The plateau indicates the individual spin features are not important for entanglement classification. We note that the highest performance achieved is 94.5\%, which remains below the performance of the the dephasing noise model, despite the dephasing model using significantly fewer data and less complex modeling.

As illustrated in both FIG.~\ref{fig:4_way_comparasion} and FIG.~\ref{fig:GT_comparasion}, the first nine features sufficiently capture the essential information for entanglement classification. Restricting the feature set to these nine parameters reduces computational costs without sacrificing accuracy, enabling efficient training and robust performance.

\section{Entanglement quantification Model}
\label{sec:quantification}

\begin{figure*}[htbp]
    \centering
    \includegraphics[width=.7\textwidth]{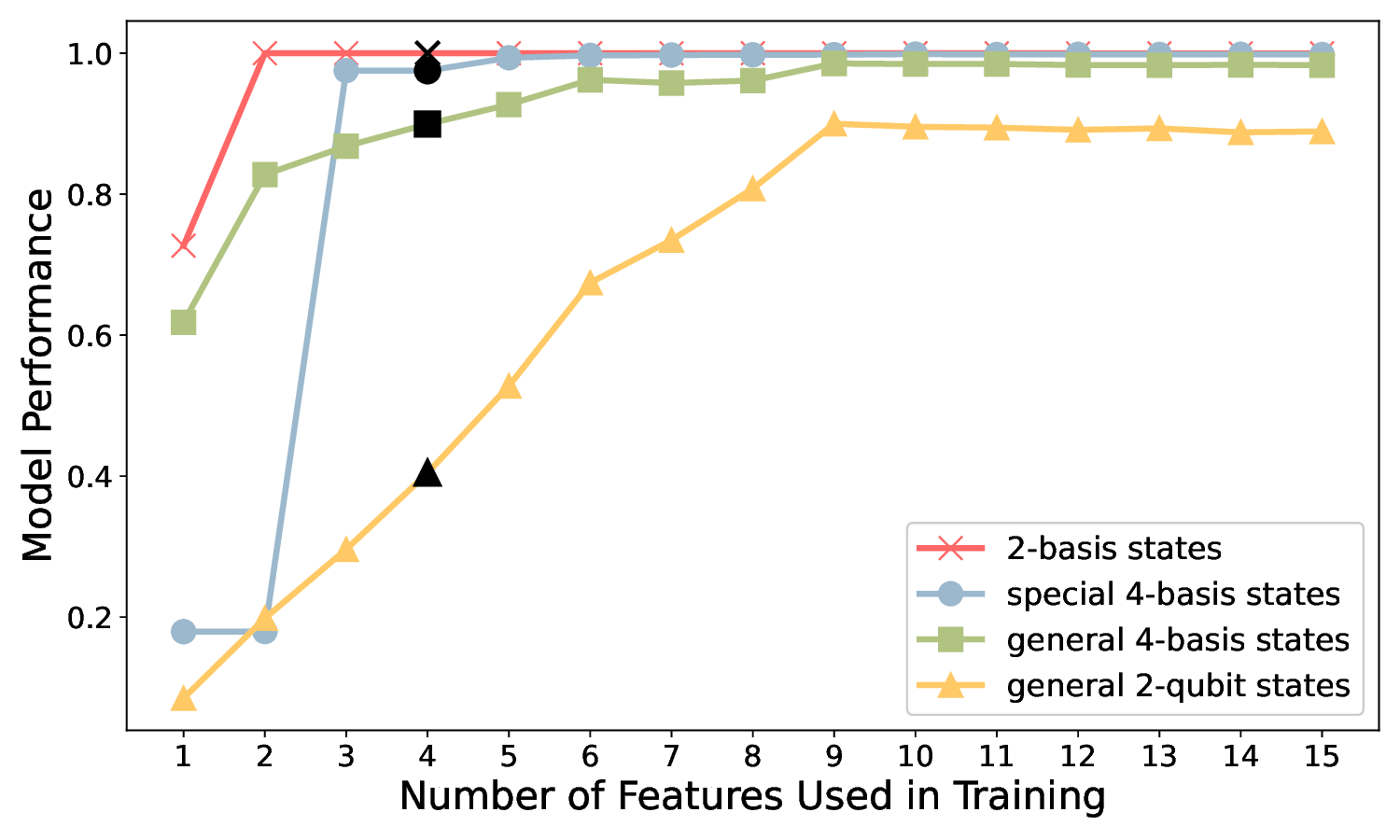}
    \caption{Model performance ($R^2$) as a function of feature number. The red line represents the special 2-basis states with dephasing noise; the blue line represents the special 4-basis states with dephasing noise; the green line denotes the general 4-basis states with dephasing noise; and the yellow line indicates the general two-qubit data. The black points indicate the performances of models trained with four Bell-like features.}
    \label{fig:concurrence_performance}
\end{figure*}

As concurrence can degrade without changing a state’s entangled or separable classification, relying solely on classification models is insufficient for describing the disentanglement process. To address this, we employ linear regression neural networks to directly predict concurrence, which quantifies the degree of entanglement in two-qubit states. The linear regression model demonstrates superior performance over classification models in certain scenarios, e.g. prediction at the entanglement boundaries. While classification models may incorrectly classify states near the entanglement boundary, the regression model consistently predicts the concurrence close to the critical value ($C = 0$). Since distinguishing between states with very low concurrence and truly separable states is often unimportant in practical contexts, the linear regression model proves particularly effective. For instance, the entangled two-qubit state such as $\sqrt{0.99999}|00\rangle+\sqrt{0.00001}|11\rangle$ (with a concurrence of $C=0.006$) can often be regarded as $|00\rangle$ in practical applications.

The following three types of pure two-qubit states experiencing dephasing noise are studied using the linear regression model. 

(1) The 2-basis states: 
\begin{align}
    |\psi_1\rangle = \cos\frac{\theta}{2}|00\rangle+\sin\frac{\theta}{2}e^{i\varphi}|11\rangle,
\end{align}
where both $\theta$, $\varphi$ parameters are randomly distributed between 0 and $2\pi$. The choices cover the entire Bloch sphere.

(2) The special 4-basis states:
\begin{align}
    |\psi_2\rangle = \cos\frac{\theta}{2}|0\rangle\left(\frac{e^{-i\varphi}|0\rangle+|1\rangle}{\sqrt{2}}\right)+\sin\frac{\theta}{2}|1\rangle\left(\frac{e^{i\varphi}|0\rangle+|1\rangle}{\sqrt{2}}\right),
\end{align}
where both $\theta$, $\varphi$ parameters are randomly distributed between 0 and $2\pi$.

(3) The general 4-basis states:
\begin{align}
    |\psi_3\rangle &= \sin\alpha|00\rangle + \cos\alpha\sin\beta e^{i\varphi_1}|01\rangle \notag\\
    &+ \cos\alpha\cos\beta\sin\gamma e^{i\varphi_2}|10\rangle + \cos\alpha\cos\beta\cos\gamma e^{i\varphi_3}|11\rangle,
\end{align}
where all parameters $\varphi_i$ ($i=1,2,3$), $\alpha$, $\beta$ and $\gamma$ are randomly distributed between 0 and $2\pi$.

After dephasing noise, the density operator for the states can be expressed as $\hat\rho_k=\sum_{i,j=1,2,3}\hat{\mathcal{E}}_{ij}|\psi_k\rangle\langle\psi_k|\hat{\mathcal{E}}_{ij}$ ($k = 1,2,3$), where $\hat{\mathcal{E}}_{ij}$ are the Kraus operators for the dephasing channel. The dephasing noise strength $p$ is randomly distributed between 0 and 1. The dataset for each type of states consists of 100,000 samples. For each state, the fifteen tomographic features are recorded, and the concurrence values are computed from Eq.~\ref{concurrence} to serve as the labels. Given the parameter distributions, the ratios of entangled states depend on the parameter distribution. The dataset for 2-basis states contains no separable instances after dephasing, while the datasets for the special and general 4-basis states have 42.1\% and 63.3\% entangled instances, respectively. Additionally, a model for the general two-qubit mixed states, containing 42.7\% entangled states, is trained for comparison.


\textbf{Model}: For all four datasets, the same neural networks are employed for training. The neural network model consists of three hidden layers with (100, 50, 50) nodes. The three hidden layers use the ReLU activation function, and the output layer uses the linear activation function. The model is trained using the SGD optimizer with a learning rate of 0.01. The batch size is set to $n_{\text{batch}}=32$, and the number of epochs is 200. The model performance is evaluated for different subsets of features. The coefficient of determination, $R^2$, defined as
\begin{align}
    R^2 = 1-\frac{(y_{\text{pred}}-y_{\text{true}})^2}{(y_{\text{true}}-\langle y_{\text{true}}\rangle)^2,\label{R_square}}
\end{align}
 is used for evaluating the model performance. For a perfect model, $R^2$ is 1. For a constant output model, $R^2$ is 0.

\subsection{Regression model performance comparison}
The performance of the linear regression model for four datasets is presented in FIG.~\ref{fig:concurrence_performance}. The qualitative behavior of the model performance is consistent with previous results in entanglement classification. Notably, no fluctuations are observed, showing that there are no under-fitting in the model.

FIG.~\ref{fig:concurrence_performance} illustrates the following salient features. All four models show improved performance with an increasing number of training features, up to nine features (all spin correlations). Beyond this point, including the last six features (single-qubit spin features) does not enhance performance. The model performance is also influenced by the complexity of the dataset, quantified by the number of free parameters used to determine the density operators. The complexity order for the four datasets is as follows:

General two-qubit mixed states (15 parameters) $>$ general 4-basis states (7 parameters) $>$ special 4-basis states (3 parameters)$=$ 2-basis states (3 parameters).

Optimal model performance is higher for datasets with less complexity, indicating that machine learning model can better capture concurrence information for simpler datasets. Additionally, for less complex datasets, optimal performance can be achieved with fewer features. For the 2- and 4-basis pure states, the first four features (Bell inequality features) suffice to predict concurrence with optimal performance. However, nine features are necessary for general pure states and general two-qubit mixed states to achieve optimal performance.

\subsection{Predicted concurrence distributions}
\begin{figure*}[htbp]
    \centering
    \includegraphics[width=\linewidth]{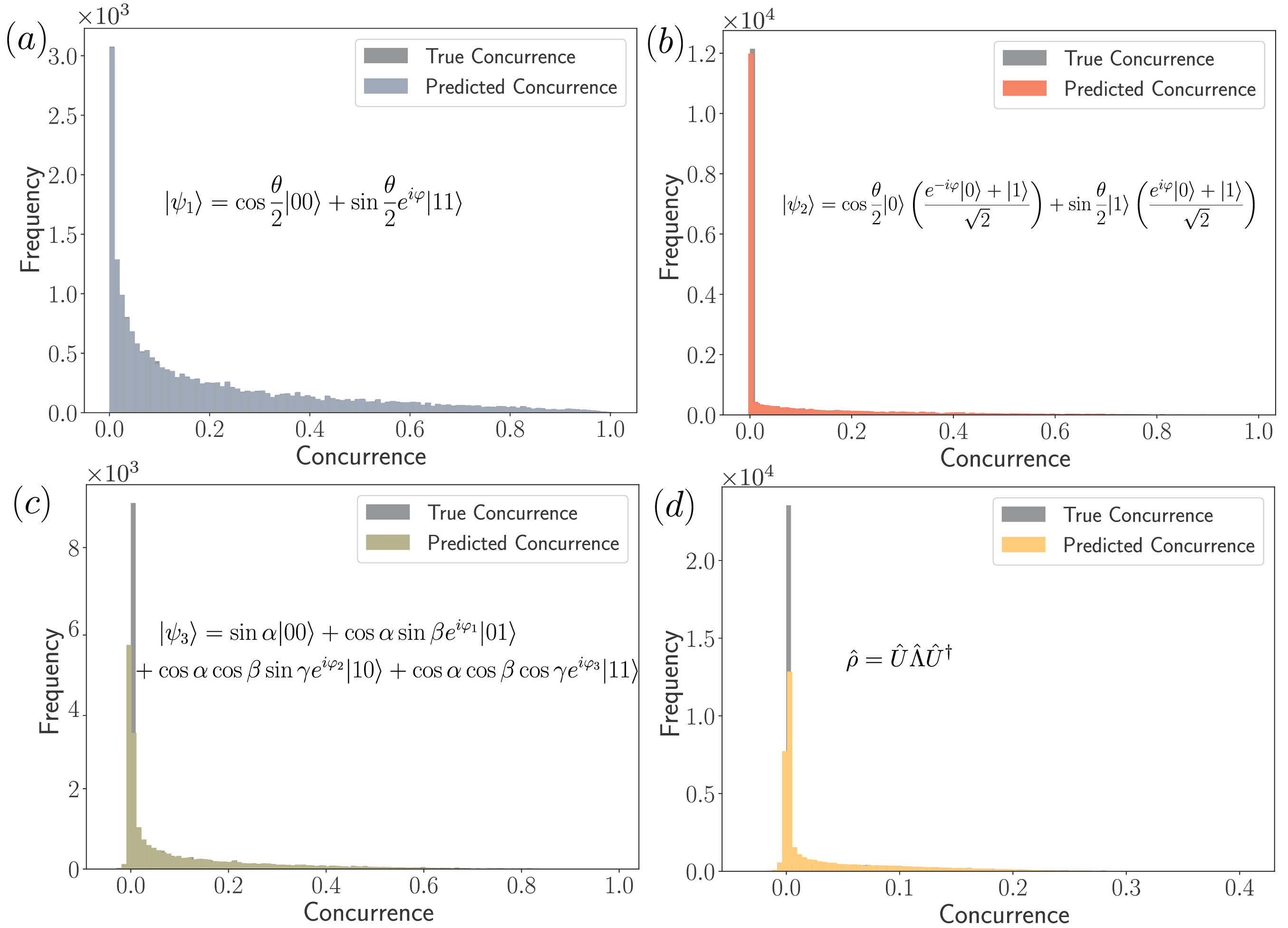}
    \caption{Predicted and true concurrence distributions across all four datasets. The neural network models for the predicted concurrence distribution are trained with the first four Bell-like features. (a) The 2-basis states with dephasing noise. (b) The special 4-basis states with dephasing noise. (c) The general 4-basis states with dephasing noise. (d) The general two-qubit mixed states.}
    \label{fig:concurrence_distribution}
\end{figure*}

To visualize model performance in entanglement quantification, the true and predicted concurrence distributions are shown in FIG.\ref{fig:concurrence_distribution}. Each model uses the first four features and all 100,000 data points. The performance of each model is given by the corresponding points (marked black) in FIG.\ref{fig:concurrence_performance}.

The model performance ($R^2$) for the two-basis states is nearly perfect (0.99999). The predicted concurrence closely matches the actual distribution, as illustrated in FIG.\ref{fig:concurrence_distribution}a. There are no separable states (zero concurrences). The states with lower concurrence values are more common, while those with higher concurrence appear less frequently. For the special 4-basis states (FIG.\ref{fig:concurrence_distribution}b), the model performance is 0.975, with about 57.9\% of the data being separable. The predicted concurrence distribution is in well agreement with the actual distribution, and the majority of data is concentrated at $C=0$. For general 4-basis states (FIG.\ref{fig:concurrence_distribution}c), the model performance is 0.899, with 36.7\% of the states being separable. Negative concurrences appear in the predicted distribution due to incorrectly predicted instances, stemming from the inherent limitations of linear regression outputs. We also tried the logistic activation function, constraining the output between 0 and 1. However, the logistic activation functions do not improve model performance. For general two-qubit mixed states (FIG.\ref{fig:concurrence_distribution}d), the model performance only reaches 0.404, and the predicted distribution deviates from the actual distribution.

The distribution of predicted concurrences supports the model's efficacy. Despite the restricted dataset size and the model complexity, the dephasing models can predict concurrence relatively well using only the four Bell-like features. For states close to the entanglement boundary, where classification models exhibit misclassification peak, the concurrences predicted by the linear regression model are close to 0, reflecting its robustness in this critical region.

\section{Discussion}

In this paper, we employ machine-learning algorithms to investigate the disentanglement processes of two-qubit quantum states subject to dephasing noise due to the environment. Our specialized ANN algorithms for dephasing noise achieve improved accuracy under restrained dataset sizes and reduced model complexity. An interesting and non-trivial question is whether this approach can also be used to investigate the disentanglement processes for multi-qubit systems. For systems with more than two qubits, there is no sufficient and necessary criterion (e.g. PPT criterion) to classify entanglement for general states. To address this, datasets can be constructed by transforming known states (e.g., product states, W states, GHZ states) using stochastic local operations assisted by classical communication (SLOCC)~\cite{LI2006}, which preserve entanglement categories. Nonetheless, constructing datasets for three-qubit states under dephasing noise remains challenging, as dephasing can alter entanglement classifications after SLOCC transformations. A possible strategy is to combine different entanglement witnesses to detect specific classes of tripartite entanglement in three-qubit states after dephasing noise. The advantages of specialized models become even more significant for neural networks trained on multi-qubit quantum states, where training data and computational resources are more constrained.
 
\begin{acknowledgments}
    This project has been made possible in part by grant number 2020-225832 from the Chan Zuckerberg Initiative DAF, an advised fund of Silicon Valley Community Foundation.
\end{acknowledgments}	


\end{document}